# Developing a Preservation Metadata Standard for Languages

Udaya Varadarajan[1,2][0000-0002-4936-0272] and Sneha Bharti[1][0000-0001-7363-9153]

[1] Documentation Research and Training Centre, Indian Statistical Institute, Bangalore

[2] Department of Library and Information Science, University of Calcutta, Kolkata

snehabharti@drtc.isibang.ac.in

**Abstract:** We have so many languages to communicate with others as humans. There are approximately 7000 languages in the world, and many are becoming extinct for a variety of reasons. In order to preserve and prevent the extinction of these languages, we need to preserve them. One way of preservation is to have a preservation metadata for languages. Metadata is data about data. Metadata is required for item description, preservation, and retrieval. There are various types of metadata, e.g., descriptive, administrative, structural, preservation, etc. After the literature study, the authors observed that there is a lack of study on the preservation metadata for language. Consequently, the purpose of this paper is to demonstrate the need for language preservation metadata. We found some archaeological metadata standards for this purpose, and after applying inclusion and exclusion criteria, we chose three archaeological metadata standards, namely: Archaeon-core, CARARE, and LIDO (Lightweight Information Describing Objects) for mapping metadata.

## 1. Introduction

Language is a means of social and emotional communication with others. Since the beginning of civilization, the human species has tried to communicate. Be it the drawings in the caves, hieroglyphics on the clay tablets or more modern - through prints and publishing. There are currently 7000 languages, of which 2500 are considered endangered languages (Crystal 2000; Krauss, 1992). Due to the global dominance of a certain language, individuals are compelled to study that language while neglecting their own. As a result, numerous cultural languages are becoming extinct day by day. Now, it is crucial to prevent the extinction of these languages. With the loss of language, the culture, people and community are also lost. Just as we preserve or conserve the endangered species of the world, language needs to be conserved or protected. The conservation and preservation of language have been studied for more than a decade (Bernard 1996; Haerudin 2018; Krylova and Renkovskaya 2020).

Several technologies, including digital repositories and archival collection management systems, can aid us in storing, preserving, and retrieving materials pertaining to these endangered languages. The preservation of these languages may facilitate the study of languages by scholars and the retrieval of language-related materials by librarians, allowing the latter to provide improved services to their patrons (Bharti and Singh 2022). It also facilitates historians to study the past of the language and the culture, curators to help in the curation and preservation of the language and government officials in policy making.

While there are archives and repositories for languages, organisations set up for endangered languages, it is found that there is significantly less study on metadata for languages or language preservation. The preservation of language when adhering to the FAIR principle (Wilkinson et al., 2016) will make it findable, accessible, interoperable, and reusable. With the aid of repositories (such as Dspace and ArchiveSpace), it is simple to locate and gain access to digital materials; however, they are not designed to facilitate the discoverability of languages. Now, in order to resolve this issue, we must develop preservation metadata that facilitates the discoverability and reuse of these endangered languages. This has been the motivating factor for the current work. The scope of the study is on the existing metadata schema for preservation. With this study, the authors aim to identify the schema that is appropriate for language preservation.

This paper is divided into 6 sections. Following the introduction is the literature study. Section 3 is the methodology adopted for the study. Section 4 details the study on metadata. The paper describes the finding in section 5 and the conclusion in section 6.

## 2. Literature Review

In the area of curation and preservation of the information science domain, there are multiple techniques and tools to protect the information, data and resources. The digitization of these languages can play a vital role in their preservation. Historically, acquiring or retaining knowledge was difficult due to the lack of available technologies. However, the introduction of technology has made it easier for us to acquire and store information. These languages could be preserved through technological means (Bharti and Singh 2022).

There are many tools and techniques of preservation. One of them is through web applications, developed for preservation (such as (Dutsova, 2016). Such software systems have two parts - 'Corpus' and 'Dictionary'. These work as lexical databases and also help in the formal representation of the language. The database is also used as a repository for language resources. The linguistic details can be stored and displayed for further use. Similarly, Dimitrova et al (2011) developed an online dictionary for Bulgarian-Polish language. The dictionary provides a two-way translation, search facility for a word in either language and allows addition and deletion. Another tool that helps is the repositories and archives such as ArchivesSpace (2022), Mukurtu (2020) and so on. Then there are means for digital archiving. The work by Berez (2013) created a comparative study of two small archives for oral traditions of language. These archives were built with the purpose of either for the academic or for the community. Through a collaborative approach, the University of Houston Libraries developed a set of 23 metadata elements in addition to the already existing element set for their digital assets (Washington and Weidner, 2017). In another collaborative work, libraries at the University of Nebraska-Lincoln (UNL) worked together to create a metadata application profile (MAP). This was done in order to record and disseminate information about the metadata standards and content procedures used by each of the four digital repositories. Platform-specific restrictions, content limitations, approaches to metadata and description, and contrasting ideologies were the main roadblocks. Through cooperative work, the group discovered similarities and decided on a minimal set of necessary metadata components for all of their repositories. After deciding on the bare minimum of metadata components, the team created and made available a LibGuide for the UNL MAP (Mering & Wintermute, 2020). The Worcester Art Museum (WAM) in order to better the preservation and retrieval of the museums' visual archive have attempted to embed the metadata with the resources (Gillis, 2016). Another project from the collaborative effort of Indiana University, University of Texas, AVP 5 and New York Public Libraries, developed an Audiovisual Metadata Platform (AMP) for metadata generation of born-digital audio and moving images (Dunn et al.,2021). Other than the works mentioned above, studies on metadata for preservation and archiving were also identified.

There are metadata for resources on the web (e.g.: Dublin core), cultural artefacts (e.g.: VRA Core, CIDOC-CRM), biodiversity (e.g.: Darwin Core, ABCD) and so on. There are metadata for data, for example, DDI, DCAT, DataCite and so on. There are various metadata schemas and standards for archaeology and heritage studies (CARARE metadata schema), historical and philosophical studies (DDI (Data Documentation Initiative) and history (OAI-ORE (Open Archives Initiative Object Reuse and Exchange). A crosswalk of the standards (either of the same domain or of the same purpose) will enable interoperability among the standards. Some examples of the mapping are MODS to various flavours of MARC such as MARC21 and MARCXML, the BIBFRAME to MARC21, the Dublin Core Metadata Elements Set and its mapping to USMARC (Caplan & Guenther, 1996). Metadata Standards Crosswalk by Getty Research Institute has the elaborate mapping of CDWA metadata schema to Dublin Core, MOD, VRA Core and CIDOC CRM.

From the literature studied, it was observed that there is a lacuna in the metadata for languages. There are many technical aids such as web or mobile applications, archives and repositories for preservation and maintenance. This work aims to fill this knowledge space by attempting to study the preservation metadata and identify the elements appropriate for the language preservation.

## 3. Methodology

A systematic approach was adopted for the study. The methodology employed to study the metadata is detailed below.

Step 1:
Search for metadata: In order to achieve the objective of the paper, we need to perform a study on the existing metadata standards. The authors searched across various academic databases such as Web of Science, SCOPUS, LISA etc. The keywords for searching were 'metadata for preservation', 'cultural heritage metadata', 'metadata for archaeology', 'preservation metadata schema'. These keywords were selected to aid in the retrieval of preservation and related domain standards. The search retrieved literature on various metadata standards. In addition, the websites of Digital Curation Centre (https://www.dcc.ac.uk/), Metadata Standards Directory (https://rd-alliance.github.io/metadata-directory/standards/) from RDA (Research Data Alliance) and such were also referred. These directories also helped in identifying metadata standards.

Step 2:
Identify: In this step, the metadata schema was identified and listed. Spreadsheet was used to tabulate the metadata details such as the name, domain, area, year of publication and year of update. There was a total of 21 metadata standards that were retrieved. Studying all 21 standards is beyond the scope of this paper. To help in the selection of metadata for study certain inclusion and exclusion criteria were formulated.

Step 3:
Inclusion and exclusion criteria: As realised in the previous step, there is a need to identify the inclusion and exclusion criteria to enable us to filter out the standards. The initial filter applied was the year of original publication. Schemas published in 2010 or later were selected for study. This reduced the standards from 21 to 6. Further study of the schema helped the authors identify the area and purpose of the schema. Schema whose purpose or domain was not in alignment with the purpose of the study (which is to identify the schema for preservation) was eliminated. The final list has three schemas.

Step 4:
Final List: Final list contains three schemas - LIDO – Lightweight Information Describing Objects, CARARE metadata schema and Archaeo-core. In order to study these metadata standards, a mapping approach was taken. For the process of mapping, there has to be a standard that acts as a touchstone. For this purpose, the Dublin Core metadata standard was chosen. This is due to the simplicity, shared understanding, larger scope and the extensibility of the schema itself. The 15 element set is compact and straightforward facilitating the discovery and description of digital resources. The semantics of DC elements are easily understandable worldwide. DC helps us in finding or describing the information because semantics of its elements are universally understood and accepted. The element set of DC though originally developed in English language, has many versions in different language i.e., Thai, French, Japanese, Greek, Spanish, etc. The 15 core element set is expandable making additional resources for resource discovery.

Step 5:
Collect the elements: Using a spreadsheet, the elements of the schema are listed. Hierarchy and nestled structure of the schema were maintained. The selected metadata standards all have elements not just to describe the artefact or the asset, but rather they record details of them as well. This complex structure was noted, but since the motivation of the work lies in the identification of preservation elements that can help preserve language, only the set of elements that describe the artefact or asset or with purpose of preservation were selected for mapping.

Step 6:

Map the elements: Following the listing of the elements, as described in the previous step, the mapping is done. The mapping is only at the level of asset or artefact or of the preservation level. The authors performed the syntactic and semantic mapping with the elements. While mapping, we get new insights and new elements that can be helpful in creating preservation metadata for languages.

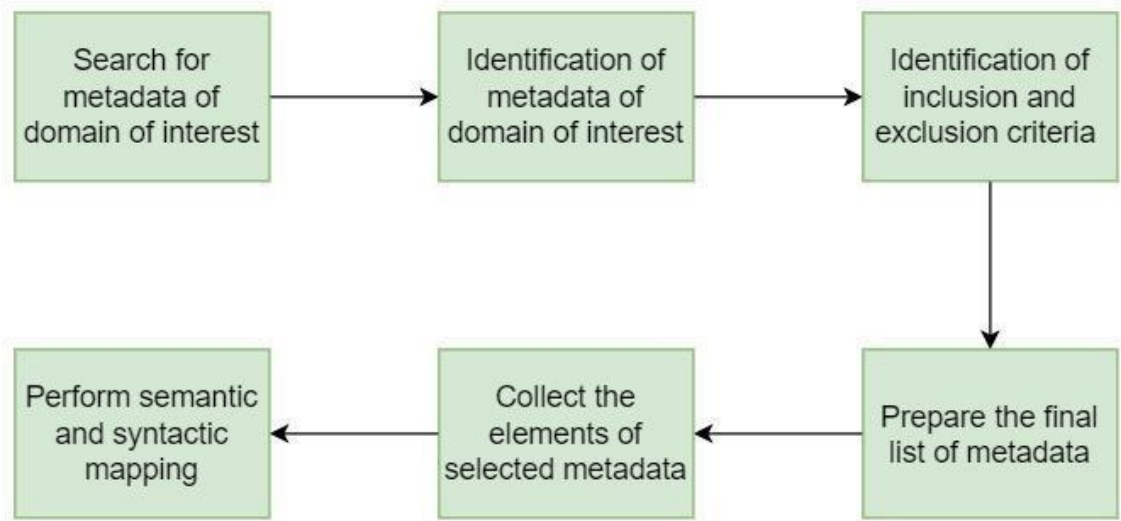

**Fig. 1.** Workflow for mapping

## 4. Mapping of Archaeological Metadata Standards with Dublin Core

The metadata schema selected for study are detailed here. The metadata schema was studied. The principles and theories of the development of these schemas were important, especially for conflict resolution. This is followed by the table that illustrates the mapping of these elements to DC. The CARARE metadata format is a harvesting schema designed to offer metadata on the online collections, historical assets, and digital resources of an organisation. The schema's strength rests in its capacity to handle the whole spectrum of descriptive information on monuments, buildings, and landscape areas, as well as their representations. The schema is an application profile based on MIDAS Heritage, a comprehensive standard designed for the complete recording of all areas of heritage management, not all of which are applicable to CARARE. The CARARE schema focuses on the precise description of historic assets, the events in which these assets were engaged, and the online locations of digital materials. The LIDO schema is designed to supply metadata for usage in a number of online services, ranging from an organisation's online collections database to portals of aggregated resources, in addition to exposing, sharing, and linking data on the web. Its strength rests in its capacity to support the whole spectrum of descriptive information for museum artefacts. It is applicable to all types of items, including art, architecture, cultural history, technological history, and natural history. The Archaeo core metadata schema is primarily for the heritage sites and the objects discovered from this place. This puts context, history and accessibility to the objects. The schema can be used in the libraries, archives and museums. The table below depicts the mapping.

**Table 1.** Mapping of Archeo Core, CARARE and LIDO to Dublin Core

| Dublin Core | Archeo Core | CARARE | LIDO |
|---|---|---|---|
| Title | Artefact Title | Appellation | Title or Object Name Set (titleSet) |
|  |  |  | Appellation Value (appellationValue) |
| alternative |  |  |  |
| Date |  |  |  |
| available |  | Designations |  |
| created | Artefact Date |  |  |
| issued | Artefact Find Date |  |  |
| modified |  |  |  |
| valid |  |  |  |
| Contributor | Artefact Photographer |  |  |
|  | Creator Role |  |  |
| Coverage |  |  |  |
| temporal | Artifact Terminus Ante Quem | Temporal |  |
|  | Artefact Terminus Post Quem |  |  |
| spatial | Artefact Spatial Coordinates | Spatial | Appellation Value (appellationValue) |
|  | Artefact Current Location |  | Location (repositoryLocation) |
|  | Artefact Origin |  | Place Name Set (namePlaceSet) |
| Creator | Artefact Creator | Actors | Appellation Value (appellationValue) |
| Description |  |  | Description/Descriptive Note (descriptiveNoteValue) |
|  | Artefact Description | Record information |  |
|  | Artefact Inscription | Inscriptions | Display State (displayState) |
|  | Artefact Condition | Description | Shape Measurements (shapeMeasurements) |
|  |  |  | Qualifier Measurements (qualifierMeasurements) |
|  | Image View Description |  |  |
| tableOfContents |  |  |  |
| abstract |  |  |  |
| Format | Artefact Form | Dimensions | Format Measurements (formatMeasurements) |
|  |  |  | Object Measurement (objectMeasurements) |
|  | Artefact Dimensions |  | Scale Measurements (scaleMeasurements) |
| extent |  |  | Extent Measurements |
| | | | (extentMeasurements) |
| | | | Display Object Measurement (displayObjectMeasurements) |
| medium | Artefact Materials | | |
| Relation | | DC:Relation | |
| hasFormat | | | |
| hasPart | | Has Part | |
| hasVersion | | | |
| isFormatOf | | | |
| isPartOf | | Is Part Of | |
| isReferencedBy | | | |
| isReplacedBy | | | |
| isRequiredBy | | | |
| isVersionOf | | | |
| references | | | |
| replaces | | | |
| requires | | | |
| Language | | | |
| Identifier | | | Description/Descriptive Note Identifier (descriptiveNoteID) |
| | Accession Number | | |
| | Artefact Munsell Number | | Place Identifier (placeID) |
| | Artefact Photograph | | Concept Identifier (conceptID) |
| | | | Legal Body ID (legalBodyID) |
| | | | Custody: Identification number (workID) |
| Publisher | Artefact Repository | Publication statement | Legal Body Name (legalBodyName) |
| Rights | | Rights | |
| Source | | | Source Appellation (sourceAppellation) |
| | | | Source Description/Descriptive Note (sourceDescriptiveNote) |
| Subject | Artefact Subject | General type | |
| | Artefact Classification | | |
| Type | Artefact Type | Conditions | |
| | | Type | |

## 5. Findings

The mapping of Archeo Core, CARARE, and LIDO to Dublin Core is shown in Table 1. Syntactic and semantic mapping has been performed for these elements. This mapping throws insights into the shared components among the three standards. Source, Publisher, Identifier, Format, Spatial, Creator, and Title are the shared components. Each standard also has elements unique to them. Though the schemas share the domain, the purpose of each standard is different and this is reflected by the elements unique to them. For instance, Artefact Techniques, Artefact Munsell Number, and Artefact Comparatives are unique elements in Archeo Core. CARARE and LIDO share unique elements. These include Provenance, Heritage Asset Type, Materials, Craft, Link, Was Present at, Is Successor to, Is Replica of, Was Digitised by, Has Representation, etc. Custody/Repository Location, Custody: Institution / Person (repositoryName), Legal Body Weblink (legalBodyWeblink), GML (gml), Term / Label (term), etc.

> Performing the mapping, it was observed that, given the need to develop preservation metadata, certain elements from these metadata standards would suffice. For example Date, Creator, Designations, Origin, Spatial, Description, Identifier, Rights, Legal Body name, Source, and Rights, among others. However, a metadata standard for language requires additional elements with focus on preservation and retrieval of languages. Added elements were identified. In order to identify the additional elements, a study was conducted on the elements of languages. This resulted in identifying elements such as script, grammatical rules, vowels, consonants, country/region, community, etc. We may also provide information on the individuals who speak that language and the linguistic family to which it belongs.

## 6. Conclusion

The purpose of the current study is to demonstrate the necessity for the development of a preservation metadata standard for endangered languages; research is ongoing. Another purpose was to identify, if any of the preservation metadata would suffice for describing and preserving languages. The establishment of language-specific metadata standards is a component of the extensive project comprising Digital Language Archiving. Our future work will concentrate on the machine-processability of the information that may be utilised as a tool for annotation, hence optimising natural language processing and machine learning.